\documentstyle[pre,aps]{revtex}

\begin{document}
\draft
\title{Derivation of continuum stochastic equations for 
discrete growth models }
\author{ Su-Chan Park and Doochul Kim }
\address{ School of Physics, Seoul National University, Seoul 151-747, Korea }
\author{ Jeong-Man Park }
\address{ Department of Physics, 
Catholic University of Korea, Puchon 420-743, Korea }
\maketitle

\begin{abstract}
We present a formalism to derive the stochastic differential equations (SDEs)
for several solid-on-solid growth models. Our formalism begins with a mapping 
of the microscopic dynamics of growth models onto the particle systems with 
reactions and diffusion. We then write the master equations for these 
corresponding particle systems and find the SDEs for the particle densities. 
Finally, by connecting the particle densities with the growth heights, 
we derive the SDEs for the height variables.
Applying this formalism to discrete growth models, we find 
the  Edwards-Wilkinson equation for the symmetric body-centered 
solid-on-solid (BCSOS) model, the Kardar-Parisi-Zhang equation
for the asymmetric BCSOS model and the generalized 
restricted solid-on-solid (RSOS) model,
and the Villain--Lai--Das Sarma equation 
for  the conserved RSOS model.
In addition to the consistent forms 
of equations
for growth models, we also obtain the coefficients associated 
with the SDEs.
\end{abstract}
\par
In recent years, the study of nonequilibrium surface growth has attracted 
considerable interest in both analytical and 
computational physics\cite{BS95}. 
A number of discrete growth models and continuum stochastic equations have 
been proposed to describe the 
kinetic roughening properties of surface growth\cite{EW82,KPZ86,KK89,VLD,KPK94,KK97}.
By studying these models and equations, 
we classify them into universality classes according to their scaling behavior
and associate the continuum stochastic equations with the given discrete 
growth models.

In general, two methods have been widely used to establish
the correspondence between a continuum 
growth equation and a discrete growth model. One method is to use Monte Carlo 
simulations to obtain the scaling exponents from the discrete model and 
compare them with those of the corresponding continuum equation. 
The other is to derive the continuum equation analytically from a given 
discrete model. Computational methodologies have contributed significantly 
to our understanding of epitaxial growth over the past few years and continue 
to do sounabated.
Analytic derivations include methods using the principle of symmetry\cite{HK92} 
or reparametrization invariance\cite{MMTB96}, and approaches starting from 
the master equation\cite{VZLW93,PK95,PK96,HG98}.
In particular, a systematic method proposed by 
Vvedensky {\it et al.}\cite{VZLW93} has been successfully applied 
to the derivation of the continuum growth equations directly from the growth 
rules of the discrete model for several solid-on-solid 
discrete models\cite{PK95,PK96,HG98}.
The derivation procedure of Vvedensky {\it et al.}
consists of two steps.
First, the discrete stochastic equation is derived for the discrete growth 
model beginning with the master-equation description of the microscopic 
dynamics of the discrete model. Second, the discrete equation is transformed 
into a continuous stochastic equation via regularization by expanding 
the nonanalytic quantities and replacing them with analytic quantities. 
In this regularization procedure, the step function is approximated by an 
analytic shifted hyperbolic tangent function, which is expanded in a Taylor 
series. As pointed out by P\v redota and Kotrla\cite{PK96}, the choice of 
regularization scheme for the step function is ambiguous. 
Thus, the coefficients in the derived continuum stochastic equation cannot 
be determined uniquely.

In this Rapid Communication, we present a method for deriving the continuum 
stochastic equations from the discrete growth models. Our method can be 
applied to most of the models that are accessible via the method of 
Vvedensky {\it et al.}
In addition to the derivation of stochastic equations consistent 
with the numerical solutions for the discrete models, our method 
predicts the coefficients in the stochastic equations.
Our method begins with 
mapping of the discrete models onto reaction-diffusion systems with 
hard-core particles and sets up the master equation of the microscopic 
dynamics in the form of the Schr\"odinger equation. 
Next, borrowing the method introduced in Ref.\cite{PKP00}, we derive the
corresponding Fokker-Planck equation, and then the stochastic
differential equation is obtained.
We apply our method 
to three discrete growth models: the body-centered solid-on-solid (BCSOS) 
model, the generalized restricted 
solid-on-solid (RSOS) model, and the conserved RSOS (CRSOS) model.

The BCSOS model is one of the simplest microscopic growth models. 
Consider a surface built from square
bricks rotated by $\pi/4$, without defects 
such as overhangs and vacancies. Each bond $i$ contains a step $S_{i}=\pm 1$. 
The growth dynamics is as follows: 
Choose one of the columns at random. If this column is at the bottom of a 
local valley ($S_{i-1}=-1$ and $S_{i}=+1$), a particle adsorbs with 
probability $p$ (and nothing happens with probability $1-p$). 
If it is at the top of a local hill ($S_{i-1}=+1$ and $S_{i}=-1$), 
a particle desorbs with probability $q$ (and nothing happens with 
probability $1-q$). 
If it is part of a local slope ($S_{i-1}=S_{i}$), nothing happens. 
This model can be mapped onto the problem of the asymmetric exclusion 
process (ASEP). The process describes particles that hop independently 
with hard-core exclusion along a one-dimensional lattice with a bias that
mimics an external driving force. 
By denoting the descending slope as the particle ($A$) and the ascending slope 
as the vacancy($\emptyset$), the microscopic growth dynamics can be mapped to a 
stochastic dynamical rule of the ASEP: the only transitions allowed for 
the site with neighboring bonds ($i-1$,$i$) are
\begin{eqnarray}
A \emptyset \rightarrow \emptyset A &~~~{\rm with~probability~}~ p,\nonumber \\
\emptyset A \rightarrow A \emptyset &~~~{\rm with~probability~}~ q.
\end{eqnarray}
This model has been studied extensively in the literature by Monte Carlo 
simulations\cite{MRSB86} and more recently it was realized that
it can be solved exactly\cite{GS92}. 
Furthermore, Derrida and Mallick calculated the diffusion 
constant associated with fluctuations of the current in the limit of 
$p \simeq q$ and derived the corresponding continuum stochastic equation with 
the coefficients correct up to the lowest order\cite{DM97}. 
Krug {\it et al.} found the coefficients for
the totally asymmetric, that is, $q=0$, case\cite{KMH92}.
Here, we apply our method to the ASEP to derive the corresponding 
stochastic equation. For simplicity, we set $p=1$
and $q=x$  ($0 \le x \le 1$). For $x=1$ the system is symmetric, whereas 
for $x=0$ it reduces to the totally asymmetric case\cite{DEM93}.

Introducing the annihilation and creation operators
($\hat a_i$'s and $\hat a_i^\dag$'s, respectively) satisfying the mixed 
commutation relations explained in Ref.\cite{PKP00}
and defining the state vector 
$|\psi;t\rangle \equiv \sum_{C} P(C;t) |C\rangle$, the master equation can be 
written as a Schr\"odinger-like equation:
\begin{equation}
\frac{ \partial}{\partial t} |\psi;t\rangle = - \hat H |\psi;t\rangle,
\end{equation}
where $P(C;t)$ is the probability for the system to be in a given 
microscopic configuration $C$ at time $t$, 
and $\hat H$, called the Hamiltonian, is an evolution operator 
expressed in terms of $\hat a_{i}$'s and $\hat a^{\dagger}_{i}$'s.
From now on, any operator will have a caret (e.g., $\hat a$,$\hat b$)
and any symbol without a caret should not be confused with
an operator. Occasionally, the same symbol is used to represent an operator and 
a density (e.g., $\hat a_i$ and $a_i$).
The Hamiltonian generating the time evolution of the BCSOS model is found 
to be
\begin{equation}
\hat H = - \sum_{i} \left( \hat a_{i}
\hat a^{\dagger}_{i+1} + x \hat a^{\dagger}_{i} \hat a_{i+1} \right).
\end{equation}
Since the diagonal terms that make $\hat H$ stochastic have no
role in our formalism, for simplicity,
we omit these terms here and throughout this Rapid Communication.
By involving the commutation relations between the Hamiltonian and some relevant
operators such as $\hat a_i^\dag \hat a_i$ and $\hat a_i^\dag \hat a_i
\hat a_j^\dag \hat a_j$, and using the property of the projection state
$\langle \cdot | (\hat a_i^\dag+ \hat a_i) = \langle \cdot |$\cite{PKP00},
where $\langle
\cdot | \equiv \sum_C \langle  C |$,
we find
the Kramers-Moyal coefficients\cite{KM} corresponding to the above
Hamiltonian:
\begin{eqnarray}
C_{i} &=& a_{i+1} + a_{i-1} - 2 a_i - (1-x) \left [
a_{i+1} ( 1 - a_i ) - a_i ( 1 - a_{i-1}) \right ],\\
C_{ij} &=& - \left [ a_i + x a_{i+1} - ( 1 - x) a_i a_{i+1} \right ]
(\delta_{i+1,j} - \delta_{ij} )\nonumber\\
&&+ \left [ a_{i-1} + x a_i - ( 1 - x) a_i a_{i-1} \right ]
(\delta_{i,j} - \delta_{i-1j} ).
\end{eqnarray}
Notice the absence of the caret on the $a$'s.
From the coefficients $C_{i}$ and $C_{ij}$, we write down the discrete 
stochastic equation
\begin{equation}
{\partial {a_{i}} \over \partial t }= C_{i} + \xi_{i},
\end{equation}
where $a_{i}$ is the local density of the particle at $i$
and $\langle \xi_{i}(t)\xi_{j}(t') \rangle=C_{ij} \delta(t-t')$. 
Replacing $a_{i}$ by $(1-\nabla h)/2$ and taking the continuum limit, 
we obtain the continuum stochastic equation for the height variable $h$
for the BCSOS model (and equivalently for the ASEP):
\begin{equation}
\frac{\partial h}{\partial t} = \frac{1-x}{2}+ \frac{1+x}{2} \nabla^{2} h - \frac{1-x}{2}
( \nabla h )^{2} + \xi(r;t),
\end{equation}
with $\langle \xi(r;t)\xi(r';t') \rangle = 4(1+x) \rho(1-\rho) \delta(r-r')
\delta(t-t')$  ($\rho$ is the stationary-state density, which is the same as
the initial density).
For the symmetric process ($x=1$), the corresponding equation is 
the EW equation, whereas for the asymmetric process ($x \neq 1$) it is 
the KPZ equation. 
(Compare the coefficients with those in Refs.\cite{DM97,KMH92}.)

Next we apply our method to the generalized RSOS growth model that was
introduced by Neergaard and den Nijs\cite{NdN97}.
They studied this model using an 
elegant mean-field type approach and derived the deterministic part
of the KPZ equation. Our method is able to produce the stochastic (noise)
part of the KPZ equation as well as the same deterministic part as the
method of Neergaard and den Nijs.
The RSOS growth model describes the growth of simple cubic surfaces 
in which only monatomic steps are allowed. The heights at the 
nearest-neighbor columns can differ by only $\Delta h = 0,\pm 1$ (the RSOS
constraint). 
After choosing one of the columns at random, one particle can be deposited 
at the site $i+1/2$ with a probability 1 
according to the height differences at $i$ and $i+1$. Employing the same 
parameter as in Ref.\cite{NdN97}, we map this model onto the two-species
hard-core particle system with the following processes:
\begin{eqnarray}
\nonumber
\emptyset \emptyset \rightarrow AB : p_h, &&\quad \quad \quad
\emptyset \emptyset \rightarrow BA : q_v,\\
\nonumber
AB \rightarrow \emptyset \emptyset : q_h, &&\quad \quad \quad
BA \rightarrow \emptyset \emptyset : p_v,\\
\nonumber
\emptyset A \rightarrow A \emptyset: p_s, &&\quad \quad \quad
\emptyset B \rightarrow B \emptyset: q_s,\\
\nonumber
A \emptyset \rightarrow \emptyset A: q_s, &&\quad \quad \quad
B \emptyset \rightarrow \emptyset B: p_s,
\end{eqnarray}
where a particle $A$ ($B$) stands for the ascending (descending) bond
and a vacuum $\emptyset$ represents the flat bond.
The corresponding Hamiltonian is found to be
\begin{eqnarray}
\hat H = - \sum_{i} \Biggl [ p_h \hat a^{\dagger}_{i} \hat b^{\dagger}_{i+1} 
+ q_v \hat b^{\dagger}_{i} \hat a^{\dagger}_{i+1} +
 q_h \hat a_i \hat b_{i+1} + p_v \hat b_i \hat a_{i+1}\nonumber\\
+p_s \hat a_i^\dag \hat a_{i+1} + q_s \hat a_i \hat a_{i+1}^\dag 
+ q_s \hat b_i^\dag \hat b_{i+1} + p_s \hat b_i \hat b_{i+1}^\dag \Biggr ] .
\end{eqnarray}
Following the same steps as above, we obtain the Kramers-Moyal coefficients 
$C_{i}^{\alpha}$ and $C_{ij}^{\alpha\beta}$ ($\alpha,\beta$ indicate either
$A$ or $B$) as follows :
\begin{eqnarray}
C^{A}_{i} &=& {p_h + q_v \over 2} v_i ( v_{i+1} + v_{i-1} ) +
{ p_h - q_v \over 2 } v_i ( v_{i+1} - v_{i-1} )
- q_h a_i b_{i+1} -p_v b_{i-1} a_i \nonumber \\
&&+{p_s + q_s \over 2} \left [ a_{i+1} + a_{i-1} - 2 a_i +
a_i ( b_{i+1} + b_{i-1} ) - b_i ( a_{i+1} + a_{i-1} ) \right ] \nonumber \\
&&+ { p_s - q_s \over 2} \left [ 
(a_{i+1} - a_{i-1} ) ( 1-2a_i ) - ( a_{i+1} - a_{i-1}) b_i - 
a_i ( b_{i+1} - b_{i-1} ) \right ],\\
C_{ij}^{AA} &=& \biggl \{ {p_h + q_v \over 2} v_i ( v_{i+1} + v_{i-1} ) +
{ p_h - q_v \over 2 } v_i ( v_{i+1} - v_{i-1} )
+ q_h a_i b_{i+1} + p_v b_{i-1}a_i \nonumber \\
&&+ {p_s + q_s \over 2} \left [
v_i ( a_{i+1} + a_{i-1} ) + a_i (  v_{i+1} + v_{i-1} )
\right ] \nonumber \\
&&+{ p_s - q_s \over 2} \left [
v_i ( a_{i+1} - a_{i-1} ) + a_i 
( v_{i-1} - v_{i+1}) \right ] \biggr \} \delta_{ij}
\nonumber \\
&&- \left \{ p_s ( 1 - a_i - b_i) a_{i+1} + q_s
a_i ( 1 - a_{i+1} - b_{i+1} ) \right \} \delta_{i+1,j}\nonumber\\
&&- \left \{  q_s ( 1 - a_i - b_i) a_{i-1} + p_s
a_i ( 1 - a_{i-1} - b_{i-1} ) \right \} \delta_{i-1,j},\\
C_{ij}^{AB} &=& C_{ji}^{BA} = \delta_{i+1,j} (p_h v_i v_{i+1} 
+ q_h a_i b_{i+1} )
+ \delta_{i-1,j} (q_v v_i v_{i-1} + p_v a_i b_{i-1} ),
\end{eqnarray}
where $v_i = 1 - a_i - b_i$ and $C_i^B$ ($C_{ij}^{BB}$) is obtained
by the exchange $a\leftrightarrow b$ followed by $i+k \leftrightarrow
i-k$ in $C_i^A$ ($C_{ij}^{AA}$).
By introducing the local slope $D\equiv a-b$ and the step density 
$S\equiv a+b$, we derive the deterministic parts of
the stochastic equations for these two parameters:       
\begin{eqnarray}
{\partial D \over \partial t} &=& 
\nabla \left [ f_q ( 1 - S )^2 + s_d ( 1 - S ) S
+ {1\over 4} h_g ( S^2 - D^2 ) \right ]\nonumber \\
&+& 
{1\over 2} s_m \nabla^2 D  + { 1\over 4} ( 2 s_m - a_s ) \nabla
\left [ D \nabla S - S \nabla D \right ],\\
{\partial S \over \partial t} &=& 2 c_s ( 1 - S )^2 - {1\over 2} a_s
(S^2 - D^2 )  + s_d \nabla \left [ ( 1 - S ) D \right ] \nonumber \\
&+& { 1\over 2} h_g [ D \nabla S - S \nabla D ] - c_s ( 1 - S ) \nabla^2 S
\nonumber \\
&+& {1\over 2} s_m \nabla^2 S - {1\over 4} a_s 
\left [S\nabla^2 S - D \nabla^2 D \right ],
\end{eqnarray}
where we have used the same notation for the parameters 
as in Ref.\cite{NdN97}; $c_s = p_h + q_v$, $a_s = p_v + q_h$, 
$s_m = p_s + q_s$, $f_g = p_h - q_v$, $h_g = p_v -q_h$, and $s_d = p_s - q_s$.
The equation for the step density $S$ is the same as that of Neergaard and den
Nijs. As pointed out in Ref.\cite{NdN97}, 
the equation for the step density contains a mass term and 
the step density reaches its stationary value $S_{0}=\left [1 + 
\sqrt{a_s / (4 c_s)} \right ]^{-1}$ after a 
characteristic time $\tau_s = (2 \sqrt{a_s c_s})^{-1}$.
Thus, although there are two order parameters, 
only the local slope fluctuates at time scales larger than $\tau_s$.  
At larger time scales, the step density does not behave as an 
independent dynamic variable. It follows local fluctuations in the 
slope of the surface:
\begin{equation}
S= S_0 + {1\over 2} \tau_s a_s D^{2} +
\tau_s \left [ s_d ( 1 - S_0 ) - {1\over 2}
h_g S_0 \right ] \nabla D + \cdots .
\label{step_slope_relation}
\end{equation}
Substituting Eq. (\ref{step_slope_relation}) for $S$ and identifying 
$D = \nabla h $, the equation for the height variable becomes
\begin{equation}
{\partial h \over \partial t} = v_{\infty} + \nu \nabla^2 h 
+ {\lambda \over 2} (\nabla h )^2  + \xi,
\end{equation}
where 
\begin{eqnarray}
v_\infty &=& f_g ( 1 - S_0)^2 + s_d ( 1 - S_0 ) S_0  + 
{1\over 4} h_g S_0^2,\nonumber\\
\nu &=&  {1 \over 4} \tau_s [ 2 s_d ( 1- S_0) - h_g S_0 ] 
[  h_g S_0 -  4 f_g ( 1 - S_0)
+ 2s_d  ( 1 - 2 S_0 )]
+ {1\over 2} s_m ( 1 - S_0) + {1\over 4} a_s 
S_0,\nonumber\\ 
\lambda
&=& -{1\over 2} h_g + {1\over 2}\tau_s a_s [  h_g S_0 -  4 f_g ( 1 - S_0)
+ 2s_d  ( 1 - 2 S_0 )],
\end{eqnarray}
and  $\langle \xi(x,t) \xi(x',t') \rangle = 
D_{\xi \xi} \delta ( x - x' ) \delta ( t - t') $ with $D_{\xi \xi}
= c_s ( 1 - S_0 )^2 + s_m  S_0 ( 1 - S_0) + a_s S_0^2/4$.
This equation is the KPZ equation corresponding to the general RSOS model.
To compare these coefficients with numerical work,
let us consider the simple RSOS model introduced by Kim and Kosterlitz 
(KK model)\cite{KK89}. The KK model corresponds to $p_h = p_v = p_s = 1$ and
$q_h = q_v = q_s = 0 $. We obtain the corresponding coefficients
$v_\infty = {4 / 9}$, $A \equiv {D_{\xi \xi} / ( 2 \nu)} = {2 / 3}$,
$\lambda = - {5 /6}$. These values are consistent with the estimated
values from a numerical study by Krug {\it et al.}\cite{KMH92}.

Recently, a different growth model with a 
RSOS condition has been proposed and studied 
by Kim et al.\cite{KPK94,KK97}. Instead of rejecting the particle when the RSOS 
condition is not satisfied, this model allows the deposited particle 
to hop to the nearest site where the RSOS condition is satisfied. 
Thus, this model has the constraint of a conserved growth condition 
and is called the conserved RSOS model. 
The detailed derivation of Villain--Lai--Das Sarma (VLD)
equation from the CRSOS model will be published 
elsewhere\cite{PPK01}.  Here we only sketch the procedure and report the
result. The procedure is similar to the one used to get the KPZ equation from
the RSOS model except for some complicated calculation. 
After some algebra, we find 
the VLD equation
\begin{equation}
{\partial h \over \partial t} = -\tilde \nu \nabla^4 h + \tilde \lambda \nabla^2
(\nabla h)^2 + \eta,
\end{equation}
where $\tilde \nu = (21 - 12 \sqrt{2})/2$, 
$\tilde \lambda = (10 - 3 \sqrt{2})/2$, and $\langle
\eta(x,t) \eta(x',t') \rangle = D_{\eta\eta} \delta(x-x') \delta(t-t')$
with $D_{\eta\eta}=(2\sqrt{2}-1)/2$.
In deriving the above equation, we kept only the most relevant terms, and
found that neither the EW nor the KPZ term exists in the growth equation.
It is known, however, that higher order terms of the form 
$\nabla (\nabla h)^{2n+1}$  $(n \ge 1)$ generate the EW term by
the dynamic renormalization group\cite{KG96}. Hence we should investigate 
the possibility of occurrence of these terms.
Indeed, we found that the dangerous term of the form $\nabla (\nabla h)^{2n +1}$
does not arise in the derivation of the VLD equation\cite{PPK01}.
Consequently, we concluded that the continuum equation of the CRSOS model
is the VLD equation.

Although the VLD equation was derived by Huang and Gu\cite{HG98} 
using the master-equation description with the regularization procedure 
proposed by Vvedensky {\it et al.}, there were some ambiguities in choosing
the regularization of the step function and thus the coefficients
could not be predicted, whereas, in our derivation, we are able to predict
the coefficients for the VLD equation corresponding to the 
CRSOS model. 
Unfortunately, however, there is no numerical method, to our knowledge, to 
find $\tilde \nu$ and $\tilde \lambda$ for a microscopic model. 
Numerical studies up to now can argue only that $\tilde \lambda$ may be 
positive\cite{KK97}, which is consistent with our derivation.

In deriving the VLD equation, we recognized the interesting aspect of the
RSOS model. When we allow hopping processes only
up to distance $l_0$, the 
stochastic equation for the height variable 
eventually becomes the KPZ one\cite{PPK01}.
This is contradictory to previously reported simulation results\cite{KY97}. 
Kim and Yook studied the RSOS model with finite-distance hopping 
by Monte Carlo simulation and concluded that there is a 
phase transition at finite $l_0$ from the KPZ class ($l_0=0$) to the
VLD class ($l_0=\infty$).
However, this conclusion seems to be a finite-size effect. 
In fact, we have confirmed  our argument
by carrying out a Monte Carlo simulation with sufficiently large system
size for several $l_0$\cite{PPK01}. 

In summary, we have presented a formalism for deriving the continuum
stochastic differential equations corresponding to discrete
growth models. Applying the formalism to the BCSOS model, we derived
the EW equation for the symmetric process and the KPZ equation with 
exact coefficients for the asymmetric process.
The RSOS model was also studied with the general probabilities
for possible processes. We derived the KPZ equation with a 
fluctuating noise part as well as the deterministic part, which is the same
as the result of Neergaard and den Nijs. For the special case with
$p$ $=$ 1 and $q$ $=$ 0 (the KK model), our coefficients are consistent
with the numerical results of Krug {\it et al.}\cite{KMH92}.
Finally, we applied our formalism to the conserved RSOS model.
For the CRSOS model,
we found that the VLD equation is the corresponding
continuum SDE. However, if we allow only finite hopping ($l_0<\infty$),
the system belongs to the KPZ
class eventually. We also predict  the coefficients of the VLD
equation for the CRSOS model.

\vspace{1cm}
This work was supported by Grant No. 2000-2-11200-002-3 from 
the BRP program of the KOSEF, and by the Brain Korea 21 Project
at Seoul National University.


\begin{references}
\bibitem{BS95} For a review, see, e.g.,
A.-L. Barab\'asi and H. E. Stanley, {\it Fractal
Concepts in Surface Growth} 
(Cambridge University Press, Cambridge, England, 1995);
T. Halpin-Healey and Y.-C. Zhang, Phys. Rep. {\bf 254},
215 (1995).

\bibitem{EW82} S. F. Edwards and D. R. Wilkinson, 
Proc. R. Soc. London, Ser. A {\bf 381}, 17 (1982).

\bibitem{KPZ86} M. Kardar, G. Parisi, and Y.-C. Zhang, 
Phys. Rev. Lett. {\bf 56}, 889 (1986).

\bibitem{KK89} J. M. Kim and J. M. Kosterlitz, 
Phys. Rev. Lett. {\bf 62}, 2289 (1989).

\bibitem{VLD} J. Villain, J. Phys. I {\bf 1}, 19 (1991); 
Z.-W. Lai and S. Das Sarma, Phys. Rev. Lett. {\bf 66}, 2348 (1991).

\bibitem{KPK94} Y. Kim, D. K. Park, and J. M. Kim, 
J. Phys. A {\bf 27}, L533 (1994)
\bibitem{KK97} Y. Kim and J. M. Kim, Phys. Rev. E {\bf 55}, 3977 (1997).

\bibitem{HK92} T. Hwa and M. Kardar, Phys. Rev. A {\bf 45}, 7002 (1992).

\bibitem{MMTB96} M. Marsili, A. Maritan, F. Toigo, and J. R. Banavar,
Rev. Mod. Phys. {\bf 68}, 963 (1996).

\bibitem{VZLW93} D. D. Vvedensky, A. Zangwill, C. N. Luse, and M. R. Wilby,
Phys. Rev. E {\bf 48}, 852 (1993).

\bibitem{PK95} K. Park and B. Kahng, Phys. Rev. E {\bf 51}, 796 (1995).

\bibitem{PK96} M. P\v redota and M. Kotrla, Phys. Rev. E {\bf 54}, 3933 (1996).

\bibitem{HG98} Z.-F. Huang and B.-L. Gu, Phys. Rev. E {\bf 57}, 4480 (1998).

\bibitem{PKP00} S.-C. Park, D. Kim, and J.-M. Park, Phys. Rev. E {\bf 62},
7642 (2000).

\bibitem{MRSB86} P. Meakin, P. Ramaulal, L. Sander, and R. C. Ball, Phys.
Rev. A {\bf 34}, 5091 (1986); P. Meakin, Phys. Rep. {\bf 235}, 189 (1993).

\bibitem{GS92} L.-H. Gwa and H. Spohn, Phys. Rev. Lett. {\bf 68},
725 (1992); Phys. Rev. A {\bf 46}, 844 (1992); D. Kim, Phys. Rev. E
{\bf 52}, 3512 (1995).


\bibitem{DM97} B. Derrida and K. Mallick, J. Phys. A {\bf 30}, 1031 (1997).

\bibitem{KMH92} J. Krug, P. Meakin, and T. Halpin-Healy, Phys. Rev. A {\bf 45}, 638 (1992).

\bibitem{DEM93} B. Derrida, M. R. Evans, and D. Mukamel, J. Phys. 
A {\bf 26}, 4911 (1993).

\bibitem{KM} See, e. g., H. Risken, {\it The Fokker-Planck Equation} 
(Springer-Verlag, Berlin, 1984); N. G. van Kampen, {\it Stochastic Processes
in Physics and Chemistry}, enlarged ed. (Elsevier, Amsterdam, 1997).

\bibitem{NdN97} J. Neergaard and M. den Nijs, J. Phys. A {\bf 30}, 1935 (1997);
there is a missing term in the expression for $\nu$ in Eq. (4.14).

\bibitem{PPK01} S.-C. Park, J.-M. Park, and D. Kim (unpublished).

\bibitem{KG96} A. K. Kshirsagar and S. V. Ghaisas, Phys. Rev. E {\bf 53},
R1325 (1996).

\bibitem{KY97} Y. Kim and S. H. Yook, J. Phys. A {\bf 30}, L449 (1997).
\end{references}
\end{document}